\documentclass[aps,prb,twocolumn,amsmath,amssymb,superscriptaddress,reprint,longbibliography]{revtex4-2}

\usepackage[utf8]{inputenc}
\usepackage{graphicx}
\usepackage{color}
\usepackage[colorlinks=true,citecolor=blue]{hyperref}
\usepackage{units}
\usepackage{orcidlink}

\renewcommand{\vec}[1]{\mathbf{#1}}

\begin{document}
	\title{Spatio-Temporal Electron Propagation Dynamics in Au/Fe/MgO(001) in nonequilibrium: Revealing Single Scattering Events and the Ballistic Limit}
	
	\author{Markus Heckschen\,\orcidlink{0000-0002-5093-5280}}
	\affiliation{Fakult\"{a}t f\"{u}r Physik, Universit\"{a}t Duisburg-Essen and CENIDE, D-47048 Duisburg, Germany}
    \author{Yasin Beyazit}
	\affiliation{Fakult\"{a}t f\"{u}r Physik, Universit\"{a}t Duisburg-Essen and CENIDE, D-47048 Duisburg, Germany}
	\author{Elaheh Shomali}
	\affiliation{Fakult\"{a}t f\"{u}r Physik, Universit\"{a}t Duisburg-Essen and CENIDE, D-47048 Duisburg, Germany}
    \author{Florian K\"{u}hne}
	\affiliation{Fakult\"{a}t f\"{u}r Physik, Universit\"{a}t Duisburg-Essen and CENIDE, D-47048 Duisburg, Germany}
	\author{Jesumony Jayabalan}
	\affiliation{Fakult\"{a}t f\"{u}r Physik, Universit\"{a}t Duisburg-Essen and CENIDE, D-47048 Duisburg, Germany}
 	\author{Ping Zhou}
	\affiliation{Fakult\"{a}t f\"{u}r Physik, Universit\"{a}t Duisburg-Essen and CENIDE, D-47048 Duisburg, Germany}
	\author{Detlef Diesing}
	\affiliation{Fakult\"{a}t f\"{u}r Chemie, Universit\"{a}t Duisburg-Essen and CENIDE, D-45711 Essen, Germany}
 	\author{Markus E. Gruner}
	\affiliation{Fakult\"{a}t f\"{u}r Physik, Universit\"{a}t Duisburg-Essen and CENIDE, D-47048 Duisburg, Germany}
	\author{Rossitza Pentcheva\,\orcidlink{0000-0002-4423-8980}}
	\affiliation{Fakult\"{a}t f\"{u}r Physik, Universit\"{a}t Duisburg-Essen and CENIDE, D-47048 Duisburg, Germany}
	\author{Axel Lorke\,\orcidlink{0000-0002-0405-7720}}
	\affiliation{Fakult\"{a}t f\"{u}r Physik, Universit\"{a}t Duisburg-Essen and CENIDE, D-47048 Duisburg, Germany}
	\author{Björn Sothmann\,\orcidlink{0000-0001-9696-9446}}
	\affiliation{Fakult\"{a}t f\"{u}r Physik, Universit\"{a}t Duisburg-Essen and CENIDE, D-47048 Duisburg, Germany}
	\author{Uwe Bovensiepen\,\orcidlink{0000-0002-1506-4491}}
	\affiliation{Fakult\"{a}t f\"{u}r Physik, Universit\"{a}t Duisburg-Essen and CENIDE, D-47048 Duisburg, Germany}\affiliation{Institute for Solid State Physics, The University of Tokyo, Kashiwa, Chiba 277-8581, Japan}

	\date{\today}
	
\begin{abstract}
Understanding the microscopic spatio-temporal dynamics of nonequilibrium charge carriers in heterosystems promises optimization of process and device design towards desired energy transfer.  Hot electron transport is governed by scattering with other electrons, defects, and bosonic excitations. Analysis of the energy dependence of scattering pathways and identification of diffusive, super-diffusive, and ballistic transport regimes are current challenges. We determine in femtosecond time-resolved two-photon photoelectron emission spectroscopy the energy-dependent change of the electron propagation time through epitaxial Au/Fe(001) heteostructures as a function of Au layer thickness for energies of 0.5 to \unit[2.0]{eV} above the Fermi energy.  We describe the laser-induced nonequilibrium electron excitation and injection across the Fe/Au interface using real-time time-dependent density functional theory and analyze the electron propagation through the Au layer by microscopic electron transport simulations. 
We identify ballistic transport of minority electrons at energies with a nascent, optically excited electron population which is determined by the combination of photon energy and the specific electronic structure of the material. At lower energy, super-diffusive transport with 1 to 4 scattering events dominates. The effective electron velocity accelerates from 0.3 to \unit[1]{nm/fs} with an increase in the Au layer thickness from 10 to 100~nm. This phenomenon is explained by electron transport that becomes preferentially aligned with the interface normal for thicker Au layers, which facilitates electron momentum / energy selection by choice of the propagation layer thickness.
\end{abstract}
	
\maketitle

\section{Introduction} \label{Sec:Intro}

Optically excited processes are widely used in energy conversion applications. In photovoltaic devices, electron-hole pairs which are excited by photon absorption are spatially separated to induce a voltage. In order to increase the photovoltaic efficiency, conversion of a single photon into more than one electron-hole pair by carrier multiplication is widely discussed \cite{green_third_2001,hanna_solar_2006}. A similar issue appears in photocatalysis and photolytic water splitting \cite{hanna_solar_2006}. An essential aspect is that optical excitations lead to
nonequilibrium phenomena. On the one hand, these conditions allow for carrier multiplication by carrier-carrier scattering since the photon energy $h\nu$ of the optical excitation is larger than the energy gap in the electronic structure. On the other hand, the nonequilibrium conditions challenge our understanding of steady-state operation. An understanding of the microscopic spatio-temporal dynamics of carrier-carrier scattering in which transport of excited carriers and their scattering with other charge carriers proceed simultaneously \cite{akel_relevance_2023} is therefore of great interest for improved energy-related applications. Properties like the ballistic mean free path of optically excited carriers can be measured in thin films by pump-probe experiments when the pump and the probe step are spatially separated  \cite{sung_long-range_2020}.

A related problem occurs in photo-induced switching of magnetization in spin valves by femtosecond laser pulses which aims at a reduced energy consumption in data storage technology \cite{igarashi_optically_2023}. The magnetic moment of a ferromagnet can be spatially redistributed in super-diffusive spin transport of optically excited carriers across interfaces \cite{battiato_superdiffusive_2010, melnikov_ultrafast_2011}. Thereby, a spin transfer torque 
can be induced \cite{razdolski_nanoscale_2017} which may lead to magnetization reversal \cite{igarashi_optically_2023} at reduced energy required compared to thermodynamic magnetization reversal. Furthermore, the combination of electronic excitations of layered systems bears further potential regarding  THz radiation generation \cite{seifert_spintronic_2022}. In such systems, electronic excitations which are induced optically in a ferromagnetic layer, the emitter, are transferred across an interface to another layer, the collector, which is electronically strongly interacting with the initially excited layer. For such phenomena, the energy-dependent scattering rate and length of optically excited carriers in heterosystems are essential because they determine the energy and momentum redistribution in these structures and have an impact on carrier multiplication.

\begin{figure}
\includegraphics[width=0.99\columnwidth]{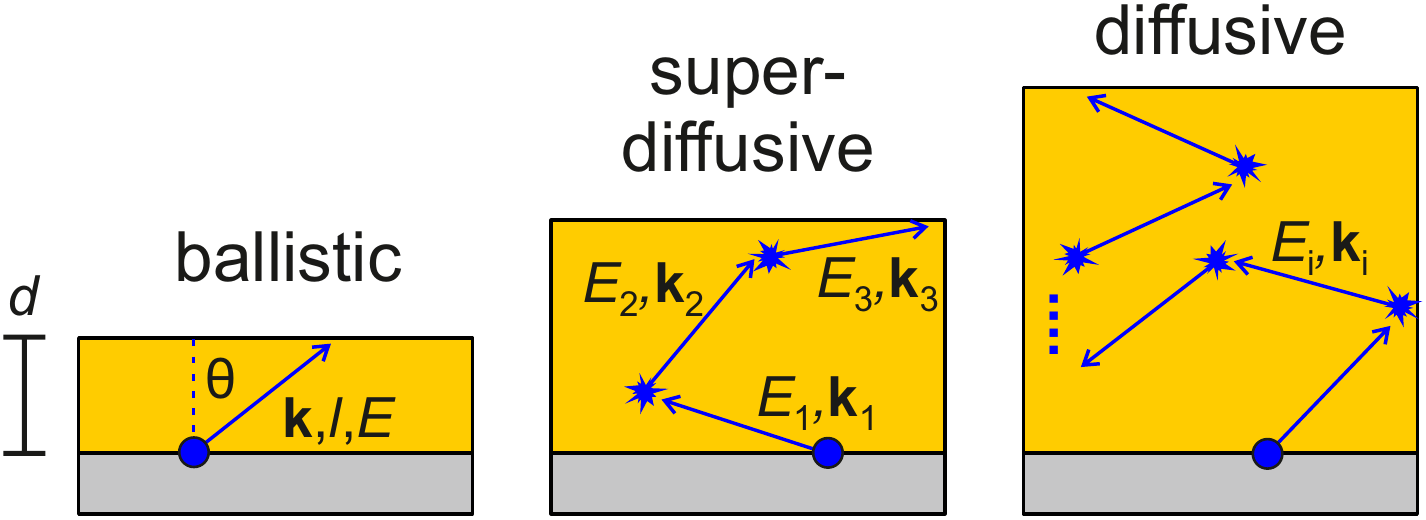}
\caption{Schematic electron scattering trajectories for ballistic, super-diffusive, and diffusive transport regimes in a heterostructure with an injection layer in grey and a propagation layer in yellow. The thickness of the latter is termed $d$. Here, $E$ is the electron energy, $\hbar\vec{k}$ the electron momentum with an injection angle $\theta$, and $l$ the length between two scattering events.}
\label{fig:fig0}
\end{figure}

Fig.~\ref{fig:fig0} depicts ballistic, super-diffusive, and diffusive transport regimes for an electron that is injected across an interface between an emitter and propagation layer. For ballistic propagation, the electron passes across the layer at constant initial energy and momentum without scattering.  In the super-diffusive regime, a small number of scattering events occurs during the electron propagation with corresponding energy and momentum changes. The secondary electrons generated in e-e scattering which ensure energy and momentum conservation are not indicated in Fig.~\ref{fig:fig0}. In the diffusive regime a very large number of scattering events occurs. Following Brownian motion the variance of a single particle distribution in space grows in thermal diffusion linearly with time, while in the ballistic regime it grows quadratically with time. Both regimes are bridged by super-diffusion in which the anomalous diffusion coefficient changes between 1 and 2 in a time-dependent fashion \cite{battiato_theory_2012}. Although Fig.~\ref{fig:fig0} suggests that with increasing thickness $d$ the transport regime changes from ballistic to diffusive, this is not the full picture, and a more detailed analysis is required. In the ballistic regime, the maximum angle $\theta$ with respect to the interface normal of the propagating electron is determined by $d=l \cos{\theta}$. Accordingly, it is the angular distribution of the propagating electrons rather than $d$ which is essential for the transport regime. For a microscopic understanding of the transport phenomena, an analysis of the number of scattering events of electrons at energy $E$ and their angular distribution under optically excited nonequilibrium conditions is desired.	

To this end, we analyze the transport and scattering dynamics of optically excited electrons in an epitaxial heterostructure MgO(001)/Fe/Au by combining \emph{ab initio} electronic structure theory  in the real-time domain and microscopic transport simulations with femtosecond time-resolved two-photon photoelectron spectroscopy. The Fe-Au heterostructure is considered here as a model system which facilitates achieving microscopic insight by avoiding complications, e.g., due to defect-induced scattering. As such, our work provides fundamental understanding of the microscopic mechanisms in nonequilibrium electron transport for a metallic heterostructure. We quantify the number of scattering events and determine the angular distribution of scattered and ballistically propagating electrons.

Excitations in solids in which electrons interact with, e.g., other electrons, the crystal lattice, and/or an ordered spin system, can be represented in the electronic bandstructure by considering energy and momentum exchange among the corresponding degrees of freedom. Measurements of charge transport properties as a function of temperature $T$ have been a leading approach to analyze these interaction processes in case of their thermal excitation. The introduction of pump-probe experiments using femtosecond laser pulses in transport studies \cite{brorson_femtosecond_1987} was an important step to measure the propagation dynamics of excitations in a time-of-flight like method but the questions of the actual scattering pathway remained a challenge. More recent femtosecond work assigned an oscillatory time-dependent response to the superposition of three ballistic contributions in momentum space close the Fermi energy \cite{liu_ballistic_2005} where the inelastic electronic scattering times are the longest according to the Fermi liquid theory.

Au/Fe/MgO(001) is a very well controlled, epitaxial heterostructure \cite{muhge_structural_1994,melnikov_ultrafast_2011,mattern_electronic_2022} with atomically sharp interfaces. We use femtosecond time-resolved two-photon photoelectron emission spectroscopy ($tr$-2PPE) \cite{petek_femtosecond_1997,weinelt_time-resolved_2002, bovensiepen_elementary_2012} as a pump-probe experiment. Following our recent development \cite{beyazit_local_2020,beyazit_ultrafast_2023}, pump and probe are spatially separated on opposite sides of the heterostructure in order to measure the effective velocity in electronic transport as a function of electron energy above the Fermi energy $E-E_{\rm{F}}$ and Au layer thickness $d_{\rm{Au}}$; see schematic in Fig.~\ref{fig:fig1}. Due to the detailed microscopic information which we showcase here, this approach will have impact on transport experiments in general. Furthermore, since Fe is ferromagnetically ordered, this development is important in the spintronic field and the spintronic THz emitters mentioned above \cite{seifert_spintronic_2022}. In comparison to previous work on spin currents \cite{malinowski_control_2008,melnikov_ultrafast_2011,bergeard_hot-electron-induced_2016} and spin filter effects \cite{alekhin_femtosecond_2017,melnikov_ultrafast_2022} our spectroscopic technique provides the energy- and angle-dependent properties. The pump-probe measurements are complemented by material-specific static and real-time time-dependent density functional theory calculations (RT-TDDFT) that provide insight in the spin-dependent excitation pattern at the Fe/Au interface and allow us to assign the main excitations peaks observed in experiment. This is coupled with microscopic electron transport simulations to determine the relaxation and propagation pathways of electrons to the Au surface.

\section{Time-resolved Two-photon Photoelectron Spectroscopy} \label{Sec:2PPE}

\subsection{Sample Preparation and Characterization}
The epitaxial Au/Fe heterostructures depicted in Fig.~\ref{fig:fig1} were grown by molecular beam epitaxy on a MgO(001) substrate. The $\unit[10\times10]{mm^2}$ MgO(001) substrates (MaTeck GmbH) were cleaned in an ultrasonic bath using ethanol, isopropanol, and acetone, subsequently. Carbon contamination of MgO(001) was removed in an Ar-O$_2$ atmosphere at $p = \unit[2\cdot10^{-3}]{mbar}$ at $T=\unit[540]{K}$. Subsequently, a \unit[7]{nm} film of Fe(001) was deposited at $T=\unit[460]{K}$ followed by deposition of Au(001) at room temperature. Pseudomorphic growth is facilitated by minimizing the lattice mismatch between MgO(001), Fe(001), and Au(001) through an in-plane rotation of the unit cells by $\pi/4$ such that MgO[010]$\|$Fe[110]$\|$Au[010] \cite{muhge_structural_1994, melnikov_ultrafast_2011, mattern_electronic_2022}. The analyzed sample was prepared as a stepped wedge with 17 steps from \unit[5]{nm} up to \unit[105]{nm} in Au film thickness $d_{\mathrm{Au}}$. Each step is \unit[400]{µm} wide to allow homogeneous pump laser excitation. The film thickness was determined by a micro quartz balance during preparation in combination with time-of-flight secondary ion mass spectroscopy of a twin sample. Scanning transmission electron microscopy was used \cite{melnikov_ultrafast_2011, razdolski_nanoscale_2017} to ensure the atomically sharp buried interface, which is crucial to avoid scattering at the interface \cite{beyazit_local_2020, beyazit_ultrafast_2023}.
	
\subsection{Experimental Setup}
\begin{figure}[t]
\centering
\includegraphics[width=0.99\columnwidth]{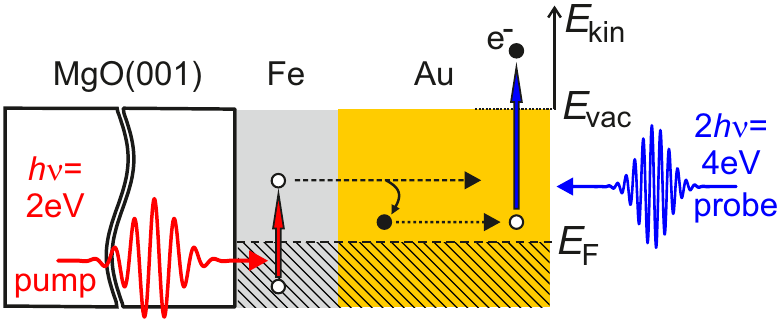}
\caption{Schematic of the $tr$-2PPE spectroscopy on Au/Fe/MgO(001). The pump pulse with a photon energy of $h\nu=\unit[2]{eV}$ excites the electrons in the buried Fe at the back of the sample which scatter during propagation through Au, are photoemitted at the Au surface by $2h\nu=\unit[4]{eV}$ photons, and analyzed in a spectrometer (not shown).}
\label{fig:fig1}
\end{figure}
	
A schematic of the {\it tr}-2PPE spectroscopy setup is shown in Fig.~\ref{fig:fig1}. The pump pulses at a photon energy $h\nu=\unit[2]{eV}$ were generated from a non-collinear optical parametric amplifier (NOPA, Clark-MXR) operated at \unit[250]{kHz} \cite{sandhofer_unoccupied_2014}. A part of the NOPA output is frequency doubled to $2h\nu=\unit[4]{eV}$ using a $\beta$-BaB$_2$O$_4$ crystal to generate probe pulses. The pulse duration of both pump and probe
pulses were determined to be less than \unit[40]{fs}. The pump and probe fluences on the sample are 50 and $\unit[1]{\mu J/cm^2}$, respectively. The pump pulse was made to enter through the MgO(001) substrate which is transparent to \unit[2]{eV} and incident on the Fe-layer. In order to ensure homogeneous excitation on a single film thickness the pump and probe beams were focused to a spot size of $\unit[140\pm40]{\mu m}$ at full width at half maximum (FWHM). Spatial overlap is determined by observing the optical transmission of the pump pulse to the Au surface on a very thin part of the Au layer and overlaying the probe beam. The angle of incidence of both the pump and probe pulses on the sample surface is 45\textdegree. The photoelectrons emitted from the sample surface are detected in a custom-built electron time-of-flight spectrometer \cite{kirchmann_time--flight_2008} in the normal direction along $z$ with an acceptance angle of $\pm$ 11\textdegree\ and analyzed regarding their kinetic energy $E_{\mathrm{kin}}$. For metallic surfaces, the maximum  $E_{\mathrm{kin}}^{\mathrm{max}}$ in 2PPE is determined by the work function $\Phi$ and $h\nu$ following $E_{\mathrm{kin}}^{\mathrm{max}}=3h\nu-\Phi$. In the present study, the intermediate electronic state is probed by a photon with energy $2h\nu$ and the energy with respect to $E_{\mathrm{F}}$ is given by
\begin{equation}
E-E_{\mathrm{F}} = E_{\mathrm{kin}}+\Phi-2h\nu.
\end{equation}
The samples were mounted in an ultra-high vacuum chamber and degassed for \unit[24]{h} at \unit[380]{K} at base pressure of $\unit[2\cdot10^{-10}]{mbar}$. All measurements were performed at room temperature (\unit[300]{K}).

The fastest temporal response of highest-energy electrons observed in the 2PPE signal was fit to a pump-probe cross-correlation. The time zero was assigned to the time delay of the maximum of this response. It is worth mentioning that because of the finite travel time of the electrons through the gold film, the determined time zero does not represent the temporal overlap of the two pulses at the pumped Fe layer, but the time delay at which the highest energy electrons have traversed the sample and reached the Au surface. Note that 2PPE is surface sensitive. We are not sensitive to the absolute travel time within the Au film but to the energy-dependent variation of this propagation time, which we term $t^*$.
	
The experimental Fe-side pumping geometry ensures that the electron is optically excited in the Fe layer and propagates through the Au layer before it is detected by photoelectron emission on the Au surface, see Ref.~\cite{beyazit_ultrafast_2023, kuhne_ultrafast_2022} for further details and comparison of the Fe-side and Au-side pumping configurations. The pronounced absorption of \unit[2]{eV} pump photons in Fe compared to pronounced reflection at this photon energy of Au facilitates injection of the spatio-temporal profile of the charge current pulse upon femtosecond laser pulse absorption in the Fe layer across the Fe/Au interface into Au. The respective optical pump absorption profiles show that more than 90\% of the pump pulse intensity is absorbed in the Fe layer, see a previous publication \cite{beyazit_ultrafast_2023}.
	
\subsection{Experimental Results}
    \begin{figure}
    \centering
    \includegraphics[width=0.9\columnwidth]{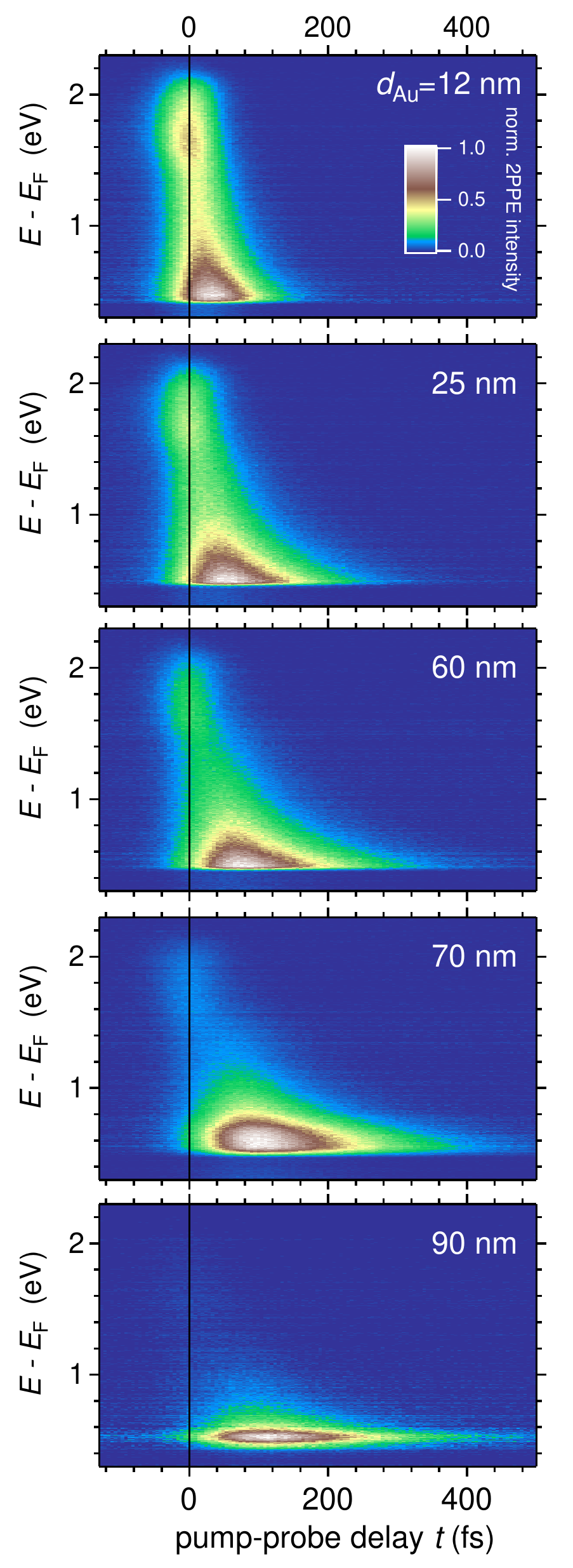}
    \caption{Time-dependent 2PPE intensity as a function of energy with respect to $E_\mathrm{F}$ in a false color representation for different Au layer thicknesses as indicated. The data were taken in the backside-pumping geometry as indicated in the sketched experimental geometry. The Fe layer thickness is kept constant at \unit[7]{nm}.}
    \label{fig:fig2}
    \end{figure}
	
\begin{figure}[t]
\centering
\includegraphics[width=0.99\columnwidth]{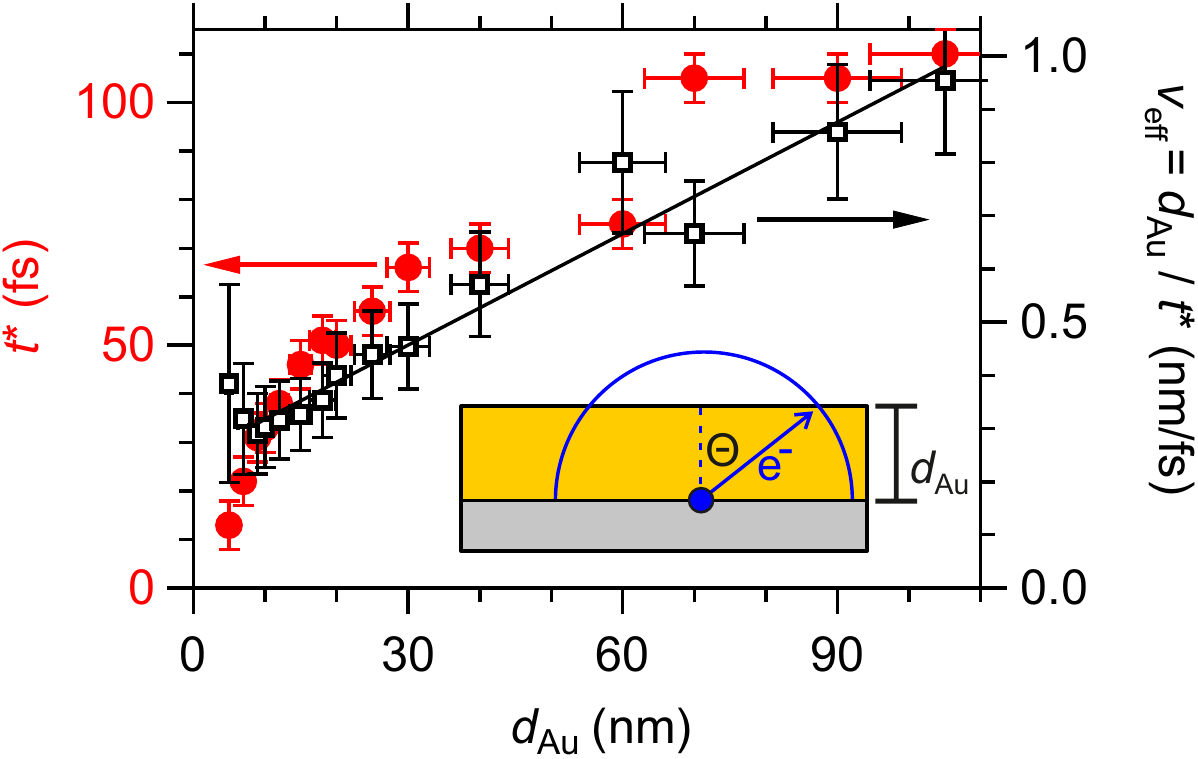}
\caption{Left axis (red label): Time delay $t^*$ due to electron propagation through Au as a function of Au layer thickness  $d_{\mathrm{Au}}$ in red solid circles for the peak maximum in 2PPE intensity determined at
$E-E_{\mathrm{F}}=\unit[0.6]{eV}$. Right axis (black label): Ratio of both quantities $t^*/d_{\mathrm{Au}}$, which represents an effective electron velocity along the interface normal direction $v_{\rm {eff}}$ as a function of $d_{\mathrm{Au}}$ in black open squares. The solid line is a linear fit to the latter data. The inset illustrates the critical angle $\theta$ for ballistic electron propagation through the Au layer upon electron injection across the Fe/Au interface. See the text for a discussion.}
    \label{fig:fig3}
    \end{figure}

In tr-2PPE, the analysis of the photoelectron intensity at a fixed energy $E-E_{\mathrm{F}}$ allows one to determine the electronic lifetime from the exponential temporal evolution \cite{petek_femtosecond_1997,bovensiepen_dynamics_2010}. This analysis was reported for the Au/Fe heterostructures previously \cite{beyazit_local_2020,beyazit_ultrafast_2023}. Here, we focus on the influence of $d_{\rm{Au}}$ on the electronic propagation through the Au layer. Figure~\ref{fig:fig2} depicts the time-resolved 2PPE intensity for the Fe-side pumping geometry as a function of energy above the Fermi energy $E_{\mathrm{F}}$ for selected $d_{\mathrm{Au}}$. The Fe layer thickness $d_{\mathrm{Fe}}$ is kept constant at \unit[7]{nm}. This value is a good compromise between the excited number of charge carriers which increases with $d_{\mathrm{Fe}}$ as long as the thickness is considerably smaller than the optical absorption depth and the scattering length of few nm which leads to a relaxation of the optical excited charge carriers towards $E_{\mathrm{F}}$ \cite{zhukov_lifetimes_2006}. Two spectroscopic signatures are observed as a function of energy $E-E_{\mathrm{F}}$ and assigned to different location of their origin based on their respective dependence on $d_{\mathrm{Au}}$. The spectroscopic signature at $E-E_{\mathrm{F}}=\unit[1.7]{eV}$ decreases in intensity relative to the intensity maximum at \unit[0.6]{eV} with increasing $d_{\mathrm{Au}}$, such that for $d_{\mathrm{Au}}=\unit[90]{nm}$ it is barely visible in Fig.~\ref{fig:fig2}, bottom. Time zero $t=0$ is determined by the time delay between the two laser pulses at which the highest energy electrons at $E-E_{\mathrm{F}}=\unit[2.2]{eV}$ are detected. The spectral signature at \unit[1.7]{eV} occurs also at $t=0$ as can be recognized in Fig.~\ref{fig:fig2}. The physical origin of this feature is an interface state which is assigned based on real-time time-dependent DFT calculations presented below in Sec.~\ref{Sec:RTTDFT}. The second spectral signature at $E-E_{\mathrm{F}}=\unit[0.6]{eV}$ near the low-energy cutoff is found to occur systematically at later time delay $t^*$ the larger $d_{\mathrm{Au}}$ is. Given the surface sensitivity of photoelectron emission spectroscopy, this behavior indicates that with larger $d_{\mathrm{Au}}$ the electron requires a longer time to propagate through the Au layer to the Au/vacuum interface. Following this hypothesis, we plot, as a function of $d_{\mathrm{Au}}$, the time delay $t^*$ at which the electrons at $E-E_{\mathrm{F}}=\unit[0.6]{eV}$ are detected at the gold surface, see Fig.~\ref{fig:fig3}.

As shown by the red data points, the dependence of $t^*$ on $d_{\mathrm{Au}}$ is sub-linear in the investigated thickness range from 5 to \unit[105]{nm}. In consequence, the ratio of $d_{\mathrm{Au}}$ and $t^*$ which represents an effective electron velocity $v_{\rm eff}$ along the interface normal direction at $E-E_{\mathrm{F}}=\unit[0.6]{eV}$, increases with $d_{\mathrm{Au}}$, as depicted in Fig.~\ref{fig:fig3}. The observed dependence $v_{\rm eff}(d_{\mathrm{Au}})$ follows a linear behavior within experimental error bars. This observation can be understood as follows; see Fig.~\ref{fig:fig3}, inset, for illustration.
Upon injection across the Fe/Au interface the electron has a wide phase space for its propagation in Au. Assuming an isotropic distribution of available momenta in Au and a ballistic propagation velocity $v_\mathrm{F}=\lambda/\tau$ electrons have equal probability to propagate the mean free path $\lambda$ in all directions in Au during the time $\tau$, as sketched by the semi circle in the inset of Fig.~\ref{fig:fig3}.
The electrons are injected with an energy of approximately \unit[1.7]{eV}. Therefore, the electrons need at least one inelastic scattering event to reach an energy of \unit[0.6]{eV} which requires a path length which is at minimum the mean free path $\lambda$. This can be connected to the thickness of the gold layer by $d_\mathrm{Au}=\lambda\cos{\theta}$, with $\theta$ as the minimum angle to ensure one inelastic collision.
The velocity component along the interface normal direction is given by
\begin{align}
	v_{\rm eff}=\frac{d_\mathrm{Au}}{t^*}=\frac{d_\mathrm{Au}}{t_1-t_0}=v_\mathrm{F}\frac{d_\mathrm{Au}}{\lambda-d_\mathrm{Au}},
\end{align}
where we assumed that the time $t_0$ is set by ballistic electrons propagating with $\theta=0$ through the sample and the time $t_1$ is the time electrons with \unit[0.6]{eV} need to reach the sample surface. Additionally, we have assumed that all electrons propagate with the same velocity $v_\mathrm{F}=\unit[1.4]{nm/fs}$ \cite{liu_ballistic_2005, mustafa_ab_2016, gall_electron_2016} which is a reasonable assumption in the observed energy range \cite{nenno_particle--cell_2018, zhukov_lifetimes_2006}. A series expansion for small gold layer thickness, $d_{\rm Au}\ll\lambda$, gives the observed linear dependence of $v_{\rm eff}$
\begin{align}
	v_{\rm eff}\approx v_\mathrm{F}\frac{d_\mathrm{Au}}{\lambda}=\frac{d_\mathrm{Au}}{\tau}.
\end{align}
A fit to the obtained linear behavior in $v_{\rm eff}(d_{\mathrm{Au}})$ results in $\tau=\unit[143\pm10]{fs}$, which is in very good agreement with literature data for inelastic hot electron lifetimes obtained by tr-2PPE \cite{bauer_hot_2015}. Since it is unlikely that for all $\theta(d_{\mathrm{Au}})$ a single scattering event occurs during the propagation across the Au layer, we investigate various scattering pathways in simulations reported in Sec.~\ref{Sec:Boltzmann} below. These account for the increasing width of the distribution in $t$ for larger $d_{\mathrm{Au}}$ obtained in Fig.~\ref{fig:fig2}.
	
\section{Real-time time-dependent Density Functional Theory} \label{Sec:RTTDFT}

To obtain a microscopic understanding of the Fe-side pump, Au-side probe procedure, we carried out RT-TDDFT calculations of a Fe$_5/$Au$_5$(001) heterostructure excited by a laser pulse with a photon energy of ${h\nu=\unit[2]{eV}}$, corresponding to the experimental pump pulse.

\subsection{Method and Details}

The periodically repeated Fe$_{5}$/Au$_{5}$(001) superlattice consists of five monolayers of bcc Fe and five monolayers of fcc Au along the (001) direction. This structure was optimized in the framework of density functional theory (DFT) with the projected augmented wave method as implemented in the VASP code \cite{kresse_efficient_1996,kresse_ultrasoft_1999} using the generalized gradient approximation (GGA) of Perdew, Burke, and Ernzerhof (PBE) \cite{perdew_generalized_1996} for the exchange correlation potential, a plane-wave cutoff of 450\,eV and a $15 \times 15 \times 5$ $k$-mesh for reciprocal space sampling.

The electronic structure and RT-TDDFT calculations in the real-time domain were performed with the full-potential linearized augmented plane wave code ELK \cite{noauthor_elk_nodate}, starting from the previously optimized geometry. For the exchange-correlation functional we have chosen the local spin density approximation (LSDA) in the parameterization of Perdew and Wang (PW92) \cite{perdew_accurate_1992}, which shows a close agreement with the VASP calculations regarding the electronic structure in the static case. Due to the presence of Au, we include spin-orbit coupling (SOC) beyond the scalar-relativistic approximation. To keep the numerical effort manageable, we  used a plane wave cut-off parameter $RK_{max}=7$ and a $k$-point grid of $8 \times 8 \times 3$, which proved sufficient in our previous investigations \cite{gruner_dynamics_2019,shomali_anisotropic_2022}.

The RT-TDDFT approach implemented in ELK propagates the electron density in time by integrating the time-dependent Kohn-Sham equations \cite{krieger_laser-induced_2015, elliott_ultrafast_2016, dewhurst_laser-induced_2018, elliott_optimal_2016, dewhurst_efficient_2016}. Time-dependent exchange and correlation were described within the adiabatic LSDA based on the PW92 functional.  The electric field of the laser pulse enters the Kohn-Sham Hamiltonian as a velocity gauge. To keep the calculations feasible, the simulated pulses are shorter than in experiment. 
To limit the pulse duration, the electromagnetic wave is folded with a Gaussian envelope with a constant full width at half maximum (FWHM) of $\unit[5.81]{fs}$, which corresponds to a finite width of $\unit[0.6]{eV}$ (FWHM) in the frequency spectrum. The maximum of the pulse with a peak power density of $S_{\rm peak}\approx \unit[5\times 10^{11}]{\rm W/cm^2}$ is reached at $t=\unit[11.6]{fs}$ after the start of the simulation. The excitation process is analyzed in terms of the layer resolved time-dependent density of states $D_{\sigma}(E,t)$, which records the transient orbital occupation numbers, projected onto their respective energies in the static, ground state density of states (DOS) \cite{dewhurst_laser-induced_2018}. This quantity has been employed recently to analyze the transient carrier dynamics obtained from a TDDFT approach applied to a Fe$_1$/(MgO)$_3$(001) heterostructure excited by laser pulses in the optical regime \cite{gruner_dynamics_2019,shomali_anisotropic_2022} and allows us to assess the relevant excitation processes and the transfer of carriers within the first $\unit[20]{fs}$, i.\,e., after the laser pulse in this calculation. 

\subsection{Results}
\begin{figure}[!htp]
\includegraphics[width=0.48\textwidth]{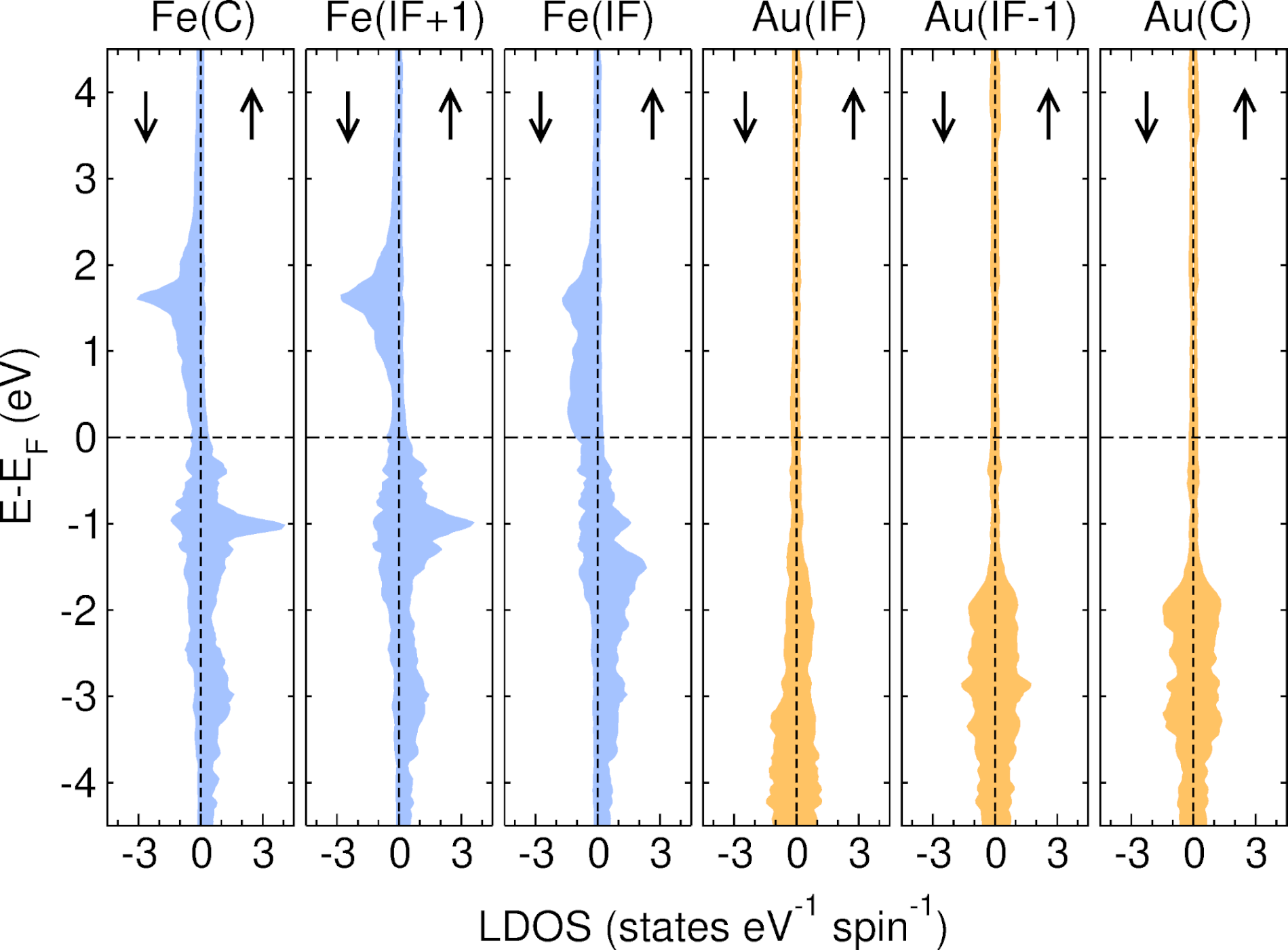}
\caption{Spin- and layer-resolved static DOS of the Fe$_{5}$/Au$_{5}$(001) heterostructure obtained with ELK. Here, (C) refers to the central layers of Fe and Au, which have the largest separation from the interface layers (IF), while (IF+1) and (IF-1) denote the intermediate layers of Fe and Au, respectively.
}
\label{FeAu-LDOS}
\end{figure}

The static layer- and spin-resolved DOS of the heterostructure is shown in Fig.\ \ref{FeAu-LDOS}. For Fe, the majority $3d$ states are largely occupied and reach up to $E_{\rm F}$, whereas the minority $3d$ band is partially occupied and extends up to \unit[2.5]{eV} above the Fermi level. We find a sharp peak at $\unit[\sim 1.7]{eV}$ in the central Fe layers and a somewhat smoother distribution in Fe(IF). In turn the unoccupied majority Fe states are dominated by $4sp$ bands. The occupied Au $5d$ band starts \unit[1.5]{eV} below $E_{\rm F}$ and shows a slight asymmetry between majority and minority spin channels, especially in the interface Au layer induced by the proximity to ferromagnetic Fe. Above $E_{\rm F}$,  Au $6s$ and $p$ states prevail, exhibiting a low DOS due to the large dispersion of these bands.  For a not too intense pulse, the ground state (i.\,e.\,static) DOS gives a first indication where relevant excitations may occur.  In the dipole approximation, a particularly large response might be expected when either the initial or the final states (or both) correspond to a region of large DOS. For the central Fe(C) layer, the latter is the case for the peak at \unit[+1.7]{eV} (above the Fermi level) which consists mainly of minority $e_g$ states, with the orbital lobes oriented along the Cartesian axes. These can be excited from $d$-states just below the Fermi level. States with $t_{2g}$ character form a broadened maximum around $\unit[-1]{eV}$ (below $E_{\rm F}$) and might be excited to the region around $\unit[+1]{eV}$ by the laser pulse. In the majority channel, $e_g$ states show a peak at $\unit[-1]{eV}$ and might thus be involved in a transition to $sp$-states around $\unit[+1]{eV}$, above the upper $d$-band edge. In Au(C), optical excitations can be expected from the upper edge of the $d$-band at about $\unit[-1.5]{eV}$ and below to $sp$-states at $\unit[\sim 0.5]{eV}$ above $E_{\rm F}$.
	
	\begin{figure}[!htp]
		\includegraphics[width=0.45\textwidth]{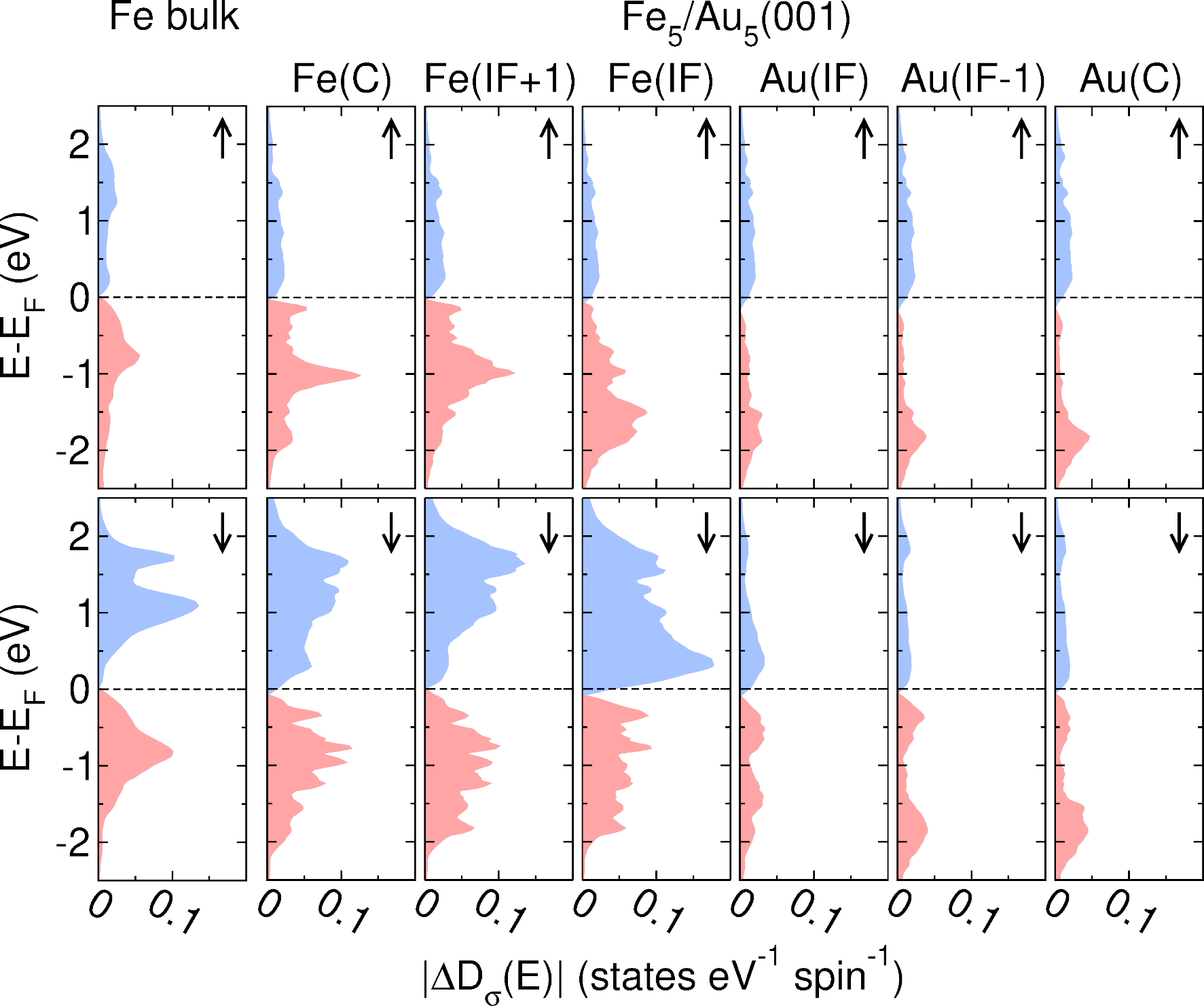}%
		\caption{The spin- and layer-resolved changes in transient occupation numbers between $t=0\,$ and $t=20.2\,$fs (after the decay of the laser pulse in this calculation), $\Delta D_{\sigma}(E)= D_{\sigma}(E,20.2{\rm fs})-D_{\sigma}(E,0)$, for in-plane laser pulse with the frequency of $h\nu=2\,$eV applied to the Fe$_{5}$/Au$_{5}$(001) heterostructure (description of layers as in Fig.\ \protect\ref{FeAu-LDOS}) and the corresponding Fe bulk material. 
        For better visibility we plot the absolute value $\left|\Delta D_{\sigma}(E,20.2{\rm fs})\right|$, the sign is indicated by the color, blue: positive sign and accumulation of occupation, red: negative sign and depletion of occupation. The upper panels refer to the majority spin channel and the lower panels to the minority spin channel.
		}
		\label{deltaDOS}
	\end{figure}

As discussed above, the layer-resolved DOS (LDOS) of both Fe(IF) and Au(IF) is modified due to the hybridization of orbitals across the interface, 
which influences the transfer of excited carriers. This is further investigated in the RT-TDDFT simulation. In this approach, the laser field acts on the entire heterostructure. In order to distinguish  whether changes in occupation result from an excitation within a specific layer in one subsystem, or whether they result from (back-)propagation of the carriers across the interface, we carried out additional RT-TDDFT calculations of the pristine bulk systems without interfaces, i.\,e.\ ferromagnetic bcc Fe and non-spinpolarized fcc Au, which were subject to the same laser pulse. The spin- and layer resolved changes in occupation of the heterostructure and the respective bulk counterparts are displayed in Fig.\ \ref{deltaDOS}. The plot shows the difference $\Delta D_{\sigma}(E)$ between the transient LDOS $D_{\sigma}(E,t)$ obtained for each spin orientation $\sigma$ at  $t=\unit[20.2]{fs}$, right after the pulse, and  at $t=0$, i.\,e., before the pulse has been applied, $\Delta D_{\sigma}(E)=D_{\sigma}(E,\unit[20]{fs})-D_{\sigma}(E,0)$. Depletion from occupied and accumulation in unoccupied states due to the laser excitation is denoted by red and blue regions, respectively.

Both for bulk Fe and the Fe layers in the heterostructure, we observe a significantly enhanced number of excited electrons in the minority channel due to the availability  of unoccupied $3d$-states above $E_{\rm F}$ in contrast to the majority $3d$ states, which lie below $E_{\rm F}$. In the minority channel of Fe(C), the excitation pattern essentially groups into three features, one at around $\unit[+1.7]{eV}$, one slightly above $\unit[+1]{eV}$ and a third, smaller one at around $\unit[+0.5]{eV}$. The excitation at $\unit[+1.7]{eV}$ is very pronounced in the minority spin channel of Fe bulk and the Fe layers in the heterostructure. It is not observed for bulk Au but we find corresponding features in Au(IF), Au(IF+1) and Au(C) of the heterostructure, see Fig.\ \ref{deltaDOS}. This means that the presence of excited carriers at this energy necessarily results from a transfer of carriers from the Fe layers across the interface. In contrast to the first two features at $+1.7\,$ and $\unit[+1.0]{eV}$, the last one at $\unit[+0.5]{eV}$ has no correspondence in the minority spin excitation spectrum of bulk Fe, but we find a corresponding signature in all Au layers of the heterostructure. This implies, that the features around $\unit[+0.5]{eV}$ and below in Fig.~\ref{deltaDOS} essentially result from hybridization of Fe and Au states at the interface. Nevertheless, since the Fe-Au interface is much thinner than 7~nm of bulk Fe and the pump light intensity is primarily absorbed in bulk Fe \cite{beyazit_ultrafast_2023}, we conclude that the electrons in 2PPE spectroscopy observed at $\unit[+0.6]{eV}$ essentially result from excitations to minority spin states in Fe bulk at higher energies and are involved in scattering processes after propagating through the Fe-Au interface. This argument is supported by the experimental observation that the 2PPE intensity at $\unit[+0.6]{eV}$ above $E_{\rm F}$ shifts to later time delay $t$ with increasing $d_{\mathrm{Au}}$, while the signature in 2PPE at \unit[1.7]{eV} above $E_{\rm F}$ only loses intensity.

\section{Transport theory} \label{Sec:Boltzmann}
\subsection{Model}
In the following, we present our theoretical modeling of the electron propagation in Au. We simulate classical, quasi-ballistic electron trajectories through a slab of thickness $d_\mathrm{Au}$ along the interface normal direction $z$ and with infinite lateral extent in $x$ and $y$. An electron with initial energy $E$ is injected at $z=0$. The initial energy is chosen randomly from a Gaussian distribution,
	\begin{equation}
		P(E)=\frac1{\sqrt{2\pi\sigma^2}}e^{-\frac{(E-\mu)^2}{2\sigma^2}},
	\end{equation}
with mean value $\mu=\unit[1.7]{eV}$ and standard deviation $\sigma=\unit[0.1]{eV}$ in agreement with the experimental observation reported in Fig.~\ref{fig:fig2} and the static and time-dependent DFT results in Figs.~\ref{FeAu-LDOS} and \ref{deltaDOS}, respectively. The direction of the initial electron propagation is chosen randomly with uniform angular distribution on the half-sphere indicated in Fig.~\ref{fig:fig3} in order to ensure propagation in the positive $z$ direction. Each electron propagates ballistically over a distance $l$ which is chosen randomly from the distribution
	\begin{equation}
		P(l)=\frac{1}{\lambda}e^{-l/\lambda},
	\end{equation}
where $\lambda$ is the energy-dependent mean free path. In the experimentally relevant energy range, electrons move to a good approximation with a constant, energy-independent ballistic velocity $v_\text{F}=\unit[1.4]{fs/nm}$~\cite{liu_ballistic_2005, mustafa_ab_2016, gall_electron_2016}. It relates the mean-free path $\lambda$ to the electron lifetime $\tau$ via $\lambda=\tau v_\text{F}$ \cite{brorson_femtosecond_1987,nenno_particle--cell_2018}. To get an analytic and realistic energy dependence of the electron lifetime and mean-free path, we parameterize the experimentally determined electron lifetime in gold~\cite{beyazit_local_2020, beyazit_ultrafast_2023, bauer_hot_2015} by
    \begin{equation}
		\tau=\frac{\tau_0}{1+\left(\frac{E-E_\text{F}}{E_0}\right)^2}, \label{eq:tau}
	\end{equation}
where $\tau_0=\unit[170]{fs}$ and $E_0=\unit[1]{eV}$ provides a good fit to the data in the considered energy range. 	This approximation captures the anomaly in the lifetime of gold at $E-E_\text{F}=\unit[1.5]{eV}$ better than Fermi liquid theory.	
	
In our model, we assume that the electron experiences an inelastic scattering event due to electron-electron interaction after having propagated for a distance $l$. The scattering event has two consequences, cf. Fig.~\ref{fig:fig0}. First, it changes the direction of propagation. The new direction is chosen randomly and is assumed to be isotropically distributed in line with the random-$k$ approximation~\cite{nenno_particle--cell_2018, battiato_ultrafast_2016}. After the scattering, the electron propagates again ballistically over a new, random distance $l$. Second, the scattering changes the electron's energy. We assume that the scattering changes the energy by a random amount $\Delta E$ which is uniformly distributed in the interval $[0,E-E_{\rm F}]$. The energy $\Delta E$ is transferred to an electron from the Fermi sea which, therefore, obtains an energy in the range $[E_{\rm F},E_{\rm F}+\Delta E]$ and whose trajectory is subsequently also included in the simulation. The secondary electron excited from the Fermi sea ensures energy and momentum conservation during the scattering process.
	
Electron trajectories are followed in the simulation until they reach the sample surface at $z=d_\mathrm{Au}$. If the electron energy drops below a threshold value of \unit[0.5]{eV}, which corresponds to the lowest energy observed in the experiment, trajectories are discarded. Furthermore, trajectories in which the electron is scattered back to the Fe/Au interface at $z=0$ are discarded as well. In each simulation sweep, we record the total path length, which is directly linked to the propagation time, the final energy, the number of collisions along the trajectory and the initial direction of propagation. In total, we have simulated $10^7$ sweeps following the propagation of the initially injected electron across the sample together with the trajectories of associated secondary electrons.
	
	\subsection{Results}
	\begin{figure}
		\includegraphics[width=\columnwidth]{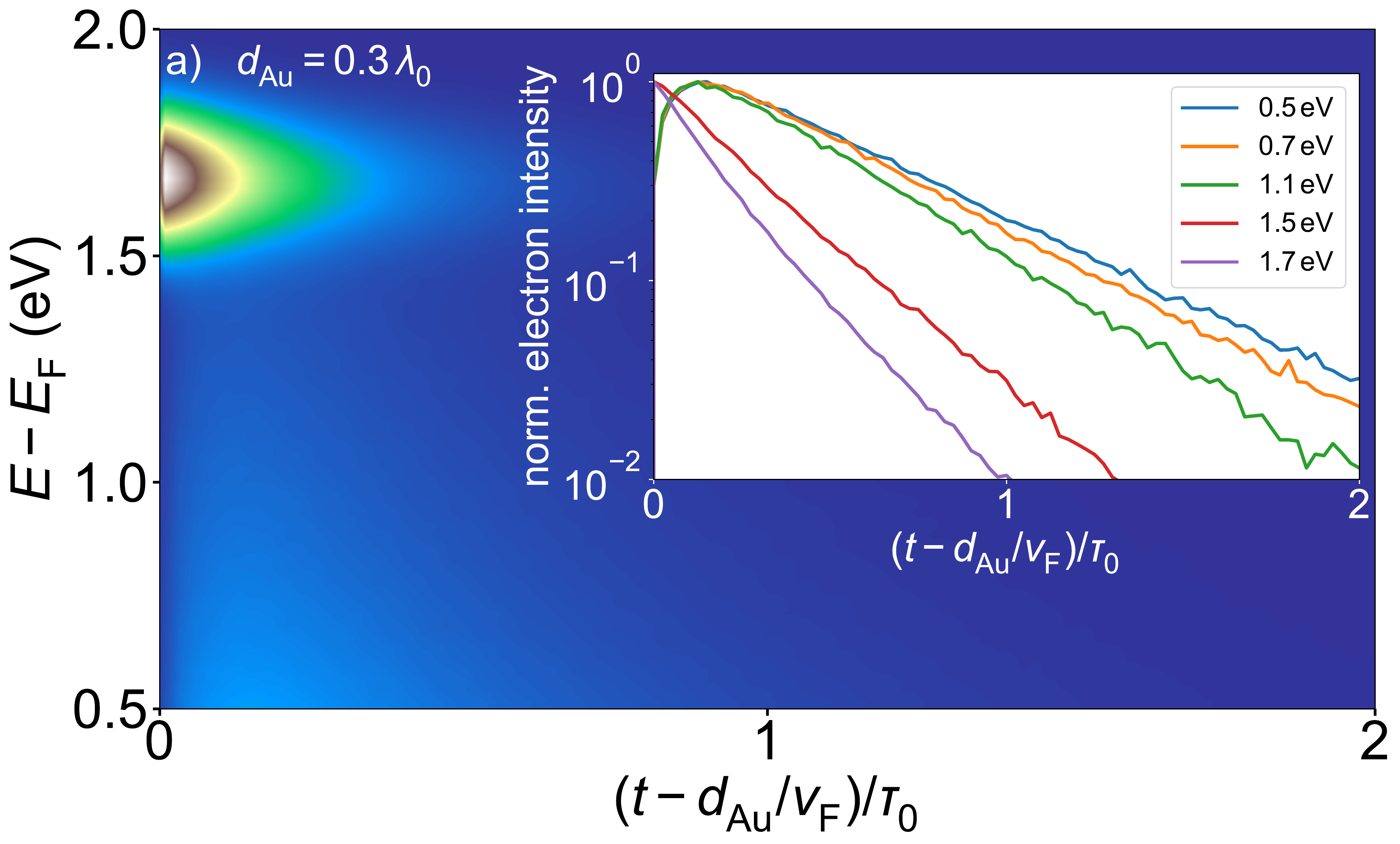}
		\includegraphics[width=\columnwidth]{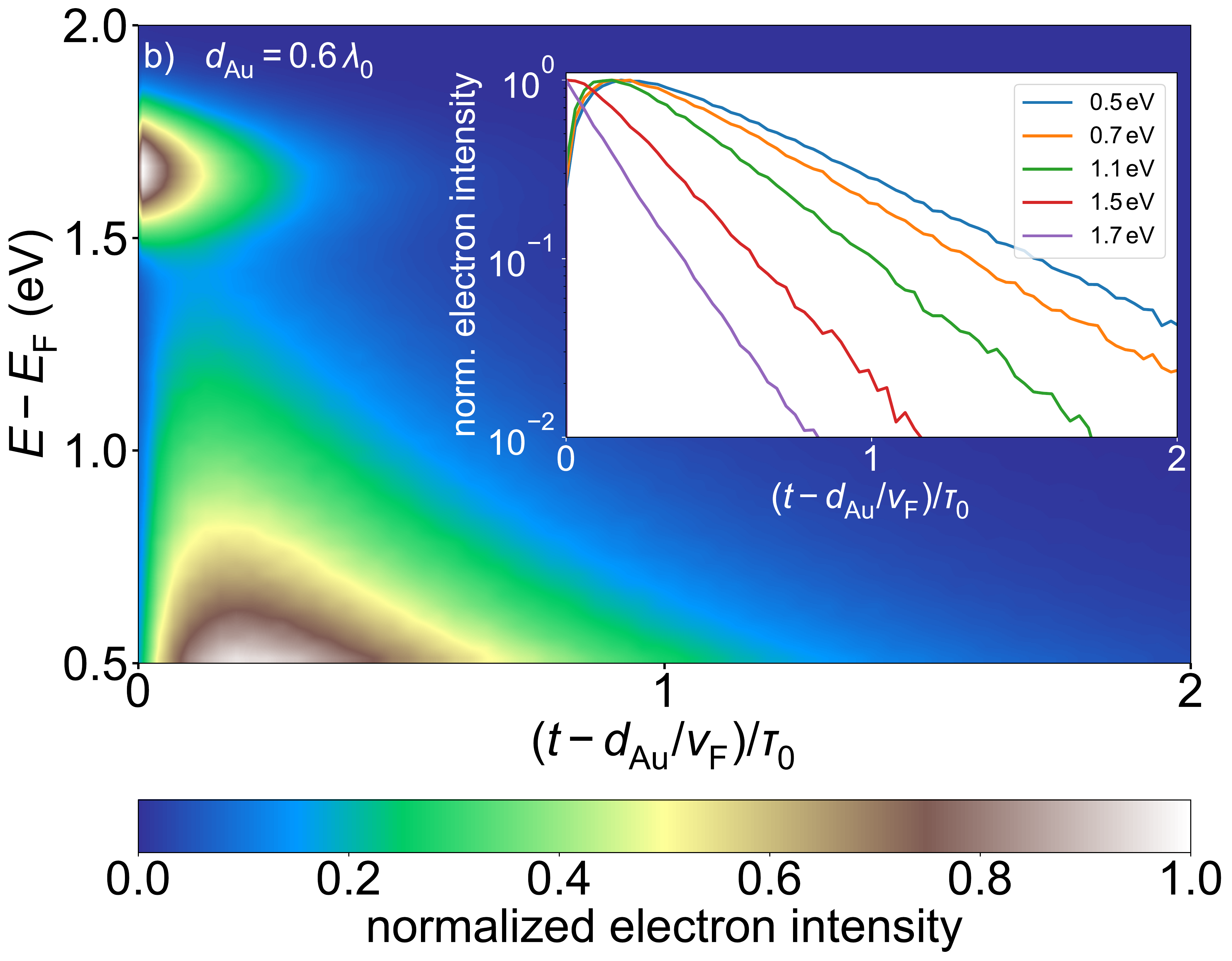}
		\caption{Normalized electron intensity as a function of energy and time for a sample thickness of a) $d_\mathrm{Au}=\unit[0.3]{\lambda_0}$ and b) $d_\mathrm{Au}=\unit[0.6]{\lambda_0}$ in a false color representation ($\lambda_0=v_\mathrm{F}\tau_0$). The insets show the time evolution of the electron intensity for depicted energies normalized to the maximum intensity of each energy.}
		\label{fig:normelecint}
	\end{figure}

In Fig.~\ref{fig:normelecint} we show the probability for electrons to arrive after a time $t$ with energy $E-E_\text{F}$ at the surface for two different sample thicknesses. In analogy to the measured tr-2PPE intensity from Fig.~\ref{fig:fig2}, we refer to this probability as the electron intensity. For both thicknesses  $d_\mathrm{Au}=\unit[0.3]{\lambda_0}$ and $d_\mathrm{Au}=\unit[0.6]{\lambda_0}$ ($\lambda_0=v_\mathrm{F}\tau_0$), we find that a significant number of electrons arrives at the surface with their original injection energy around \unit[1.7]{eV} at a time $t=d_\mathrm{Au}/v_\text{F}$. This shows essentially that these electrons propagate ballistically, i.e. without any scattering through the sample in a direction close to the $z$ axis. Here, we highlight the qualitative agreement with the measured intensities from Fig.~\ref{fig:fig2}. The corresponding electron intensity decays exponentially in time, cf. insets in Fig.~\ref{fig:normelecint}, with a relaxation time that depends on energy according to Eq.~\eqref{eq:tau} following our previous results \cite{beyazit_local_2020,beyazit_ultrafast_2023}. In principle, elastic scattering could lead to similar results. It is excluded here due to lacking agreement for experimental and simulation results, as discussed further below in Sec.~\ref{Sec:Discussion}.

For thin samples, cf. main panel Fig.~\ref{fig:normelecint}(a), the intensity at energies below the injection energy is small. At such energies, electrons must have undergone at least one inelastic scattering process which is unlikely for samples that are much thinner than the mean-free path. In contrast, for thicker samples as shown in the main panel of Fig.~\ref{fig:normelecint}(b) there is a significant electron intensity at low energies. Inelastic scattering events are more likely to happen in this situation because the electrons propagate a longer distance through the sample. The intensity at low energies shows a different time dependence compared to the high-energy electrons. As seen in the insets of Fig.~\ref{fig:normelecint} for energies $E-E_\text{F}\leq\unit[1.5]{eV}$, the intensity builds up, reaches a maximum whose position depends on the sample thickness and electron energy, and finally decays exponentially. In agreement with the experimental results, we find that the most likely transit time (maxima in the inset of Fig.~\ref{fig:normelecint}) increases sublinearly with $d_\mathrm{Au}$ indicating that the effective electron velocity increases with increasing sample thickness.
	
\begin{figure}
		\includegraphics[width=\columnwidth]{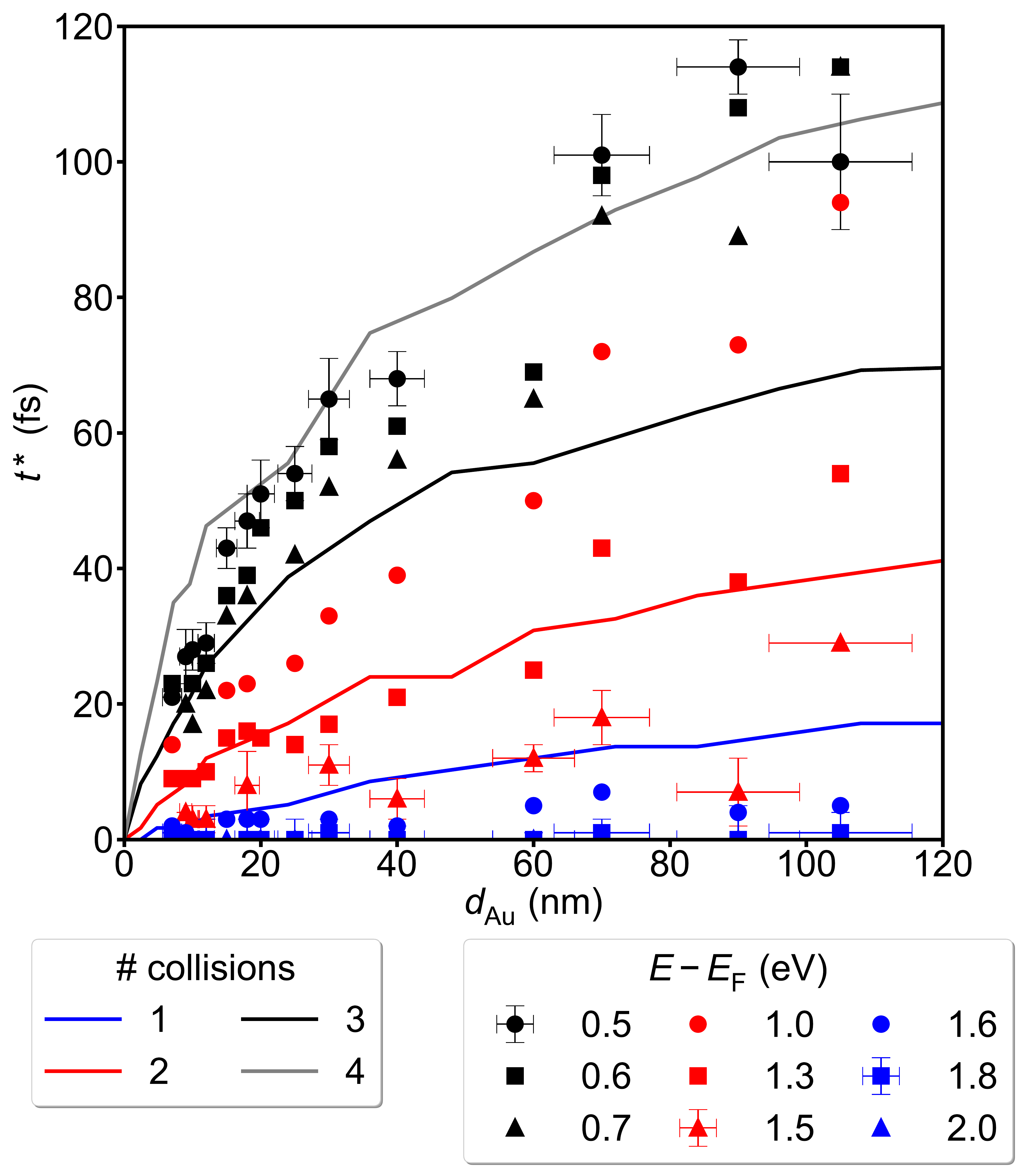}
		\caption{Comparison of the experimentally measured time delay $t^*$ for different $E-E_{\rm{F}}$ as a function of $d_\mathrm{Au}$ and the simulated trajectories with a fixed number of collisions. The simulated results are scaled to the SI unit system assuming a ballistic velocity $v_\mathrm{F}=\unit[1.4]{nm/fs}$ and $\tau_0=\unit[170]{fs}$. For a better visibility, only representive error bars are given for each group.}
		\label{fig:VeloCwExp}
\end{figure}

To understand the effective acceleration of electrons, reported in Fig.~\ref{fig:fig3}, we consider electron trajectories with a given number of inelastic scattering events separately. The time delay $t^*$ between the arrival of the ballistic high-energy electrons at $t=d_\mathrm{Au}/v_\text{F}$ and the maximal intensity of the low-energy electrons is shown in Fig.~\ref{fig:VeloCwExp} by solid lines as a function of $d_\mathrm{Au}$ for trajectories that undergo a given number of inelastic collisions. The different curves all pass through the trivial point $t^*=0$ for $d_\mathrm{Au}=0$. For thin gold layers, $t^*$ increases linearly with the sample thickness but changes to a sublinear increase above a certain threshold thickness, which increases with the number of inelastic scattering events. At the same time, $t^*$ is the larger the higher the number of inelastic collisions. From this result, we estimate that for collision numbers much larger than 4, $t^*$ will exhibit the linear growth up to much larger gold thicknesses. Assuming that this linear behavior $t^* \propto d_\mathrm{Au}$ continues up to a very large number of collisions, we estimate a propagation velocity of about \unit[0.2]{nm/fs}, i.e. seven times smaller than the Fermi velocity. In this sense, the mentioned threshold thickness represents the transition from a diffusive to a super-diffusive transport regime.

\section{Discussion} \label{Sec:Discussion}

With the obtained experimental and theoretical results we are able to provide a comprehensive microscopic picture of the optical excitation in Fe, electron injection across the interface to Au, and propagation through the Au layer. The photo-induced changes in the electronic population from the Fe bulk to the interface calculated by RT-TDDFT, see Fig.~\ref{deltaDOS}, highlight that \unit[2]{eV} photons predominantly excite minority spin electrons. These calculations also show that the excited electronic DOS of bulk Fe is essentially recovered within 3~ML distance from the interface. Since one (001) atomic monolayer (ML) is \unit[0.144]{nm} and $d_\mathrm{Fe}=\unit[7]{nm}=\unit[49]{ML}$, we conclude that the \unit[2]{eV} photons are absorbed in Fe bulk with high probability and are, according to the population gradient across the interface, injected into the Au layer via a coherent hybrid interface state within the mean free path of Fe \cite{zhukov_lifetimes_2006}. Direct optical excitation of such interface states plays a minor role. The calculated excited DOS of minority spins in Au exhibits near the interface a spectral signature with a peak at $E-E_{\rm{F}}=\unit[1.7]{eV}$ and a maximum at \unit[0.5]{eV} in excellent agreement with the experimental observation in Fig.~\ref{fig:fig2}. This agreement confirms our description of the electron injection across the interface. We note that in comparison to previous work \cite{alekhin_femtosecond_2017} which used \unit[1.5]{eV} photons and concluded injection of majority spin electrons, the RT-TDDFT results indicate that the higher photon energy used here results in a strong injection in the minority spin channel. This comparison is also interesting regarding minority electrons excited in Fe bulk to \unit[1]{eV} above $E_{\rm{F}}$. Both studies, the present one and Ref.~\cite{alekhin_femtosecond_2017}, find that these excited electrons are injected with a low probability from Fe to Au. 

The key experimental finding of a detection of electrons at the Au surface with the lower energy of $E-E_{\rm{F}}=\unit[0.5]{eV}$ which exhibit a time delay $t^*$ compared to the electron injection at \unit[1.7]{eV} motivated the description of inelastically scattered electrons by microscopic transport simulations. This description agrees very well with the experimental results, cf. Figs.~\ref{fig:fig2} and \ref{fig:normelecint}. The observation of $t^*$ in experiment and in the transport simulations provides a direct comparison of these results which connects the number of scattering events with the electron energy and the film thickness. Both the simulated and the experimentally observed $t^*(d_{\rm{Au}})$ exhibit a sublinear relation which was assigned to deviation from diffusive transport in Sec.~\ref{Sec:Boltzmann}. This consistent behavior allows one to link the electron dynamics at different energies to the number of scattering events. In Fig.~\ref{fig:VeloCwExp}, we depict these results by grouping the experimental data into lower (black), intermediate (red), and higher energy (blue). These groups follow the sub-linear relation. At the lower energy, the corresponding $t^*$ fall between 4 and 3, at intermediate energy between 3 and 1, and at higher energy below 1 scattering event. This latter result showcases the ballistic transport regime of the primarily injected electrons for $d_{\rm{Au}}<\unit[110]{nm}$. The lower energy electrons experience up to 4 scattering events, but do not reach the diffusive regime. This finding is in agreement with our previous study that analyzed electronic transport at lower energy near $E_{\rm{F}}$ in a similar experimental geometry \cite{kuhne_ultrafast_2022}, which allowed to reach the diffusion regime. Regarding ballistic electrons, our finding differs from previous work \cite{liu_ballistic_2005}, which reported ballistic contributions at particular momenta near $E_{\rm{F}}$ while we have identified ballistic electrons exclusively at the highest energy. Potentially, this difference originates from a different sensitivity on the ballistic transport in the two different experimental methods. It is possible that the larger signal contribution of scattered electrons masks a smaller ballistic contribution, as discussed in Ref.~\cite{liu_ballistic_2005}.

	\begin{figure}
		\includegraphics[width=\columnwidth]{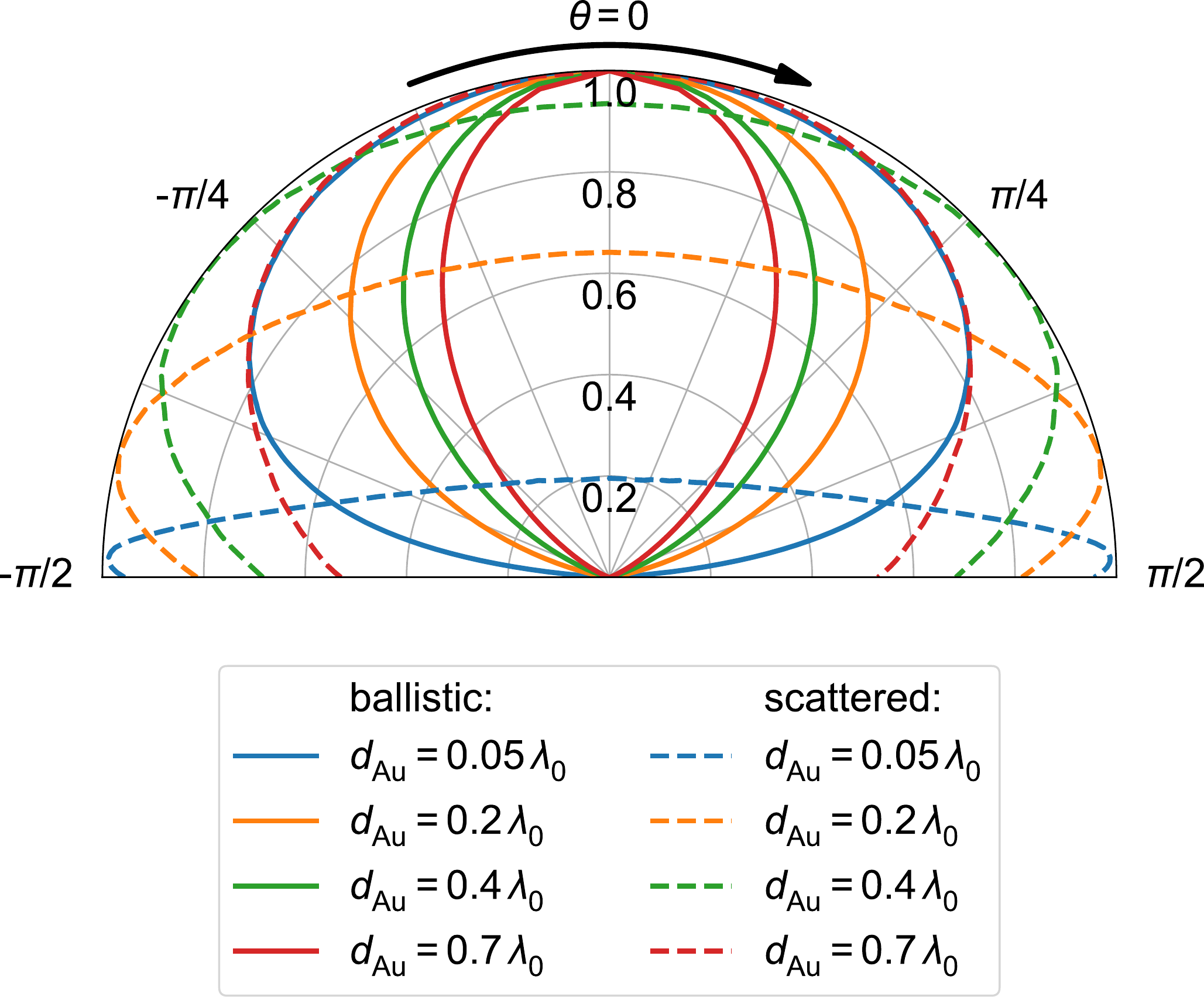}
		\caption{Normalized angular distribution of initial azimuthal angles for scattered and ballistic trajectories for sample thicknesses $d_\mathrm{Au}=\unit[0.05]{\lambda_0}$, $\unit[0.2]{\lambda_0}$, $\unit[0.4]{\lambda_0}$ and $\unit[0.7]{\lambda_0}$ with $\lambda_0=v_\mathrm{F}\tau_0$. An initial angle of $\theta=0$ corresponds to propagation parallel to the $z$ axis.}
		\label{fig:AngDis}
	\end{figure}

From the above analysis, we conclude that electrons observed at energies below \unit[1.5]{eV} experience at least one inelastic collision on average. This implies that they must propagate a minimum distance of the inelastic mean free path $\lambda$. Therefore, scattered and ballistic electrons are expected to exhibit different distributions of injection angles $\theta$ as long as $d_{\rm{Au}}<\lambda$. To confirm this, we analyze the electron momenta of the individual trajectories in our simulation. Thereby, we obtain a microscopic understanding of the transport and relaxation dynamics as well as the acceleration. In Fig.~\ref{fig:AngDis}, we show the distribution of injection angles $\theta$ for ballistic and scattered electrons for four different $d_{\rm{Au}}$. Ballistic electrons are most likely to propagate along the interface normal direction because at $|\theta|>0$ the pathway and the scattering probability increase. The thinner the Au film, the broader the distribution of initial angles becomes. For $d_{\rm{Au}}$ close to $\lambda_0$ the angular distribution of $\theta$ for ballistic electrons becomes a narrow cone. This is in agreement with the picture discussed in the inset of Fig.~\ref{fig:fig3}. In contrast, scattered electrons need a path long enough to ensure at least one collision. In sufficiently thin Au films, see $d_{\rm{Au}}=\unit[0.05]{\lambda_0}$ in Fig.~\ref{fig:AngDis}, most electrons that scatter propagate with an initial angle close to $\theta=\pm\pi/2$ almost along the interface plane. Consequently, the effective electron velocity to traverse the layer is much smaller than the ballistic electron velocity. Increasing the sample thickness shifts the most likely initial angle towards the interface normal direction, see Fig.~\ref{fig:AngDis}. Hence, the effective velocity of the scattered electrons increases for larger $d_{\rm{Au}}$ in agreement with the experimental observation in Figs.~\ref{fig:fig3} and \ref{fig:VeloCwExp}. This behavior suggests to apply momentum-dependent scattering at buried interfaces for energy selection in electron (or hole) injection across interfaces. A potential design implies electron transmission of high energy electrons along the interface normal and (total) reflection of electrons at larger angles $\theta$.

Up to now, we have only included inelastic scattering in our considerations and found the average number of inelastic collisions necessary to achieve a certain energy. We now turn to the influence of elastic scattering. Assuming an energy-independent elastic scattering time which is comparable to the inelastic lifetime of the \unit[2.0]{eV} hot electrons~\cite{nenno_particle--cell_2018}, the measured acceleration, cf.~Fig.~\ref{fig:VeloCwExp}, would change drastically. While at \unit[2.0]{eV} the ratio of elastic and inelastic scattering events would be 1:1, the ratio changes continuously to 4:1 at \unit[0.5]{eV}. Hence, the total number of scattering events would be much larger than four for the low energy electrons if elastic scattering  was present. In such a scenario, the time delay $t^*$ of the low energy electrons would increase linearly in Fig.~\ref{fig:VeloCwExp} in the considered range. Since the data points from the experiment are far below this limit, the experimental results indicate that the elastic scattering lifetime is much larger than the inelastic lifetime. Consequently, the elastic contributions are negligible in our analysis. Electronic transport and relaxation in the optically excited nonequilibrium conditions are governed by inelastic scattering events. Recalling the single crystalline sample quality, this conclusion is in good agreement with absence of defects in our heterostructure samples.
	
Finally, we address the question how to transfer the results reported here for Fe(100)/Au(100) to other material systems. It is obvious that defects in the propagation layer would add elastic scattering, see above, and limit ballistic propagation. This would hamper the angular selection along the interface normal for ballistic electrons reported in Fig.~\ref{fig:AngDis}. Defects at the interface, however, are potentially a very interesting modification opportunity. Since defects represent a source of momentum, the injection probability across the interface could be enhanced by lifting the requirement to inject via a coherent interface state to ensure energy and momentum conservation. One might ask how a change from metallic to semiconducting constituents affects the behavior. Essentially we expect that the spin and angular/momentum selection remain effective also for semiconducting systems. What would change are the scattering probabilities since e-e scattering plays a minor role in semiconductors and becomes relevant only for much higher optical excitation densities reaching a quasi chemical potential in the conduction band. Under the excitation conditions investigated here, the e-e scattering rate in the metallic case would, more or less, be replaced by e-ph scattering for a semiconducting system.	

\section{Conclusions}
In a comprehensive experimental and theoretical study, we presented a real-time analysis of nonequilibrium optical and electronic excitations in an epitaxial Au/Fe/MgO(100) heterostructure with the focus on spatio-temporal electronic transport across the layer stack. The experimental findings are based on femtosecond time-resolved two-photon photoelectron spectroscopy in which the excitation and detection are spatially separated and confined to the Fe layer and the Au surface, respectively. Thereby, we (i) measured the energy relaxation dynamics in the transport process and (ii) analyzed the transport mechanism microscopically. The microscopic understanding of the experimental data was rendered by two theoretical approaches. On one hand, RT-TDDFT allowed us to identify the electronic states involved in the excitation in the Fe part of the heterostructure and the injection across the Fe-Au interface. This microscopic insight facilitated, on the other hand, to set up a transport simulation describing the propagation and relaxation processes of the hot electrons in the gold layer. The experiment  revealed an effectively accelerated transport behavior with increasing Au layer thickness which is explained in the transport simulations by an interplay of ballistic and superdiffusive trajectories with a preferential ballistic transport along the interface normal direction. The electrons with the highest energy propagate ballistically while lower-energy electrons need to propagate super-diffusively to ensure a relaxation beforehand.

We propose to exploit this angular-dependent transport as an energy filter to select ballistic over scattered electrons, or \emph{vice versa}. The microscopic and general understanding of the transport regimes and relaxation processes on the ultrafast timescale in our model system might pave the way towards an increase in energy efficiency in devices with photo-excited charge carriers such as solar cells. Besides, our insights might be useful for other phenomena like transport in heterostructures and spin currents.

\begin{acknowledgments}
Funding by the Deutsche Forschungsgemeinschaft (DFG, German Research Foundation) within Project ID No. 278162697-SFB 1242 and through Project No. BO1823/12—FOR 5249 (QUAST) and computational time at the supercomputer magnitUDE, Center for Computational Sciences and Simulation of the University of  Duisburg-Essen (DFG grants INST 20876/209-1  FUGG, INST 20876/243-1 FUGG) are gratefully acknowledged.
\end{acknowledgments}
	
%

\end{document}